\documentclass[11pt]{article}

\usepackage{amssymb}
\usepackage[pdftex]{color,graphicx}
\usepackage[paperwidth=5.3in, paperheight=7.5in, top=.55in, bottom=.5in, left=.14in, right=.14in]{geometry}
\usepackage{textcomp}

\newtheorem{theorem}{Theorem}
\newtheorem{lemma}{Lemma}

\newcommand{\ipar}{\begin{list}{\hbox{$\:$}}{} \item[] }
\newcommand{\rapi}{\end{list}}

\begin{document}

\title{\bf The Edge Group Coloring Problem \\
with Applications to Multicast Switching}
 
\author{Jonathan Turner}
\date{\normalsize WUCSE-2015-02}
\maketitle

\begin{abstract}
This paper introduces a natural generalization of the classical edge coloring problem
in graphs that provides a useful abstraction for two well-known problems in multicast 
switching. We show that the problem is {\sl NP}-hard and evaluate the performance 
of several approximation algorithms, both analytically and experimentally. 
We find that for random $\chi$-colorable graphs,
the number of colors used by the best algorithms falls within a small constant factor of $\chi$,
where the constant factor is mainly a function of the ratio of the number of outputs to inputs.
When this ratio is less than 10, the best algorithms produces solutions that use fewer than $2\chi$ colors.
In addition, one of the algorithms studied finds high quality approximate solutions for
any graph with high probability, where the probability of a low quality solution is
a function only of the random choices made by the algorithm.
\end{abstract}

\pagestyle{plain}

\begin{figure}[h]
\centerline{\includegraphics[width=1.75in]{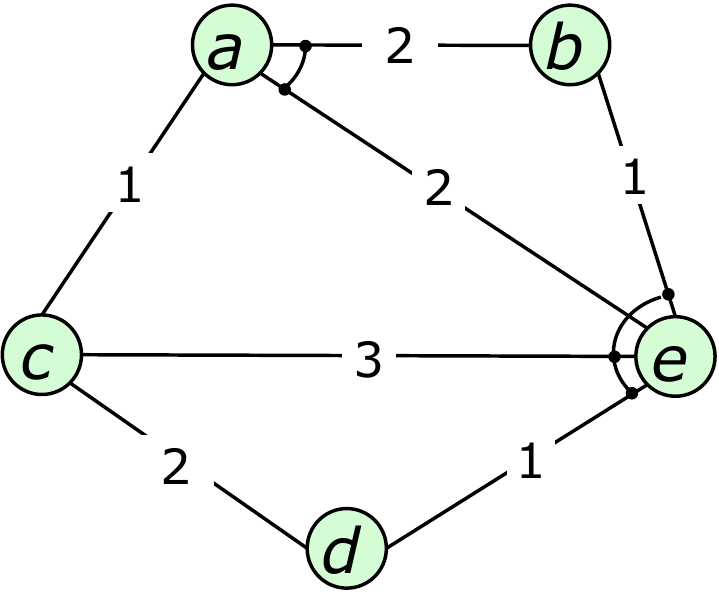}}
\caption{Example of edge group coloring}
\label{example1}
\end{figure}

An instance of the {\sl edge group coloring problem} is an undirected graph
$G=(V,E)$ and a partition of its edges into groups $\{g_1,\ldots,g_k\}$,
where the edges forming each group all share a common endpoint.
A {\sl coloring} is a function $c$, from the edges to the positive integers,
which assigns different values to pairs of edges that share a common
endpoint and belong to {\sl different} edge groups.
The objective of the problem is to find a coloring that uses
the smallest possible number of distinct colors.
An example is shown in Figure~\ref{example1};
here, the arcs joining selected edges define the groups
(so for example, one group consists of the edges $\{b,e\}, \{c,e\}$ and $\{d,e\}$).
The integers on the edges form a valid assignment of colors.

\begin{figure}[h]
\centerline{\includegraphics[width=1.25in]{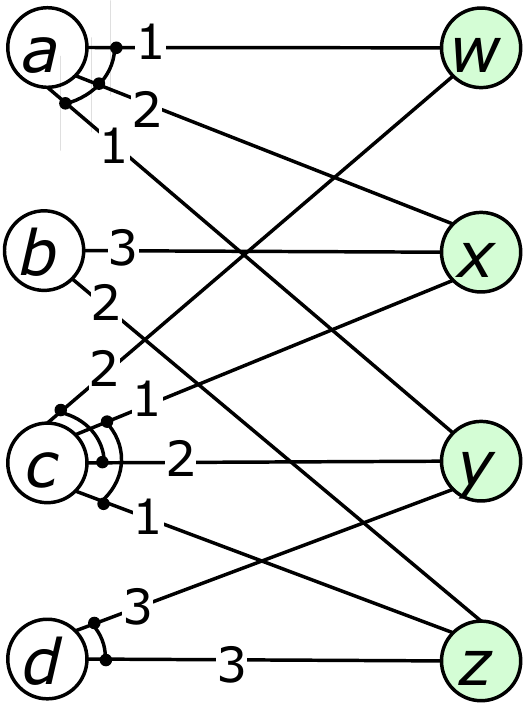}}
\caption{Restricted version of edge group coloring problem}
\label{example2}
\end{figure}

In this paper, we focus on a restricted version of the problem
in which the graph is bipartite, with the vertices divided between
{\sl inputs} and {\sl outputs}, and all groups are ``centered''
at an input.
An example of such a graph appears in Figure~\ref{example2}.
This version of edge group coloring can be applied to a version of the
multicast packet scheduling problem in crossbar switches~\cite{PMA97}.
In this application, each edge-group represents
a multicast packet, with each group including an edge for each output
that is to receive a copy. The colors assigned to edges correspond to
time-slots during which copies of multicast packets are transferred to outputs.
Copies of a multicast packets can be transferred to multiple outputs at the same time,
but there is no requirement that all copies be transferred at the same
time. This corresponds to the edge coloring rule that allows edges in the
same group to share a color.

The problem can also be applied to routing mulitcast connections in a three stage Clos network~\cite{CC53}, and indeed the multicast routing problem has been studied
extensively. This paper draws directly on two papers in the multicast routing literature.
The paper by Yang and Masson~\cite{YM91} includes a multicast routing algorithm that
can be adapted to the edge-coloring problem and provides the best worst-case
approximation bound that is currently known for edge-group coloring.
Kirkpatrick, Klawe and Pippenger~\cite{KKP85} formulate the problem of routing
connections 
in a Clos network as a hypergraph coloring problem and give bounds on the number of 
colors required. Our formulation of the problem is equivalent to theirs, 
but is expressed in somewhat simpler language. 

We start by showing that the edge-group coloring probelm is {\sl NP}-complete and give a
simple approximation method that can be refined into several specific algorithms.
These are evaluated experimentally on random $k$-colorable graphs.
In section 4, we re-visit Yang and Masson's algorithm in the context of the edge-group 
coloring problem, provide a simplified analysis of its worst-case performance and 
evaluate its performance experimentally.
In section 5, we discuss an algorithmic approach that is implicit in the
proof of the main theorem of~\cite{KKP85}. 
We give an explicit statement of this approach and consider two specific algorithms 
based on it. We show that one of these can be viewed as a randomized approximation
algorithm with a  performance ratio that is somewhat better than
the one in~\cite{YM91}.

\section{Preliminaries}

For the bipartite version of the problem studied here, we find it convenient to represent edges
as ordered pairs $(u,v)$ where $u$ is an input and $v$ is an output.
For any vertex $u$, let $\delta(u)$ denote the number of edges incident to $u$
(the vertex degree)
and let $d(u)$ denote the number of distinct groups with an edge incident
to $u$ (the {\sl group count}).
In general, $d(u)\leq \delta(u)$ and for the restricted graphs considered here,
$d(v)=\delta(v)$ for outputs $v$.
We also let $\Delta=\max_u \delta(u)$ and $D=\max_u d(u)$.
In addition, we let $\Delta_o$ be the largest vertex degree among the outputs,
and $D_i$ be the maximum group count among the inputs.

\begin{figure}[h]
\centerline{\includegraphics[width=2.5in]{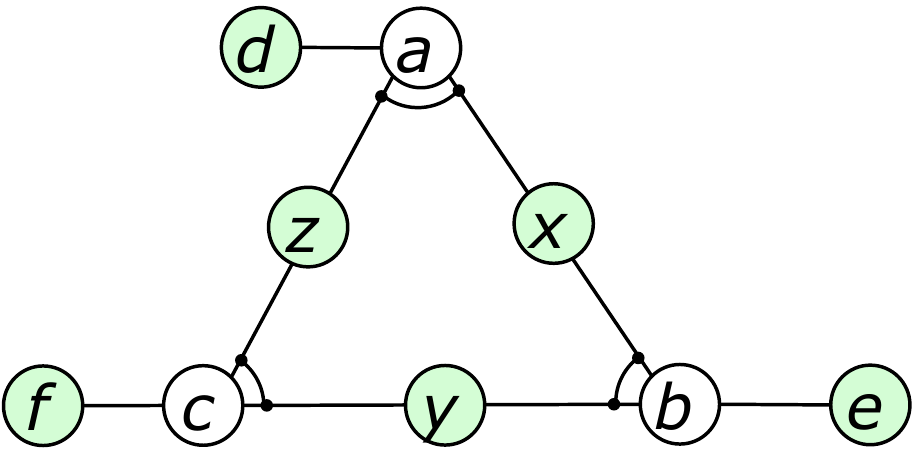}}
\caption{$D$ colors are not always enough}
\label{example3}
\end{figure}

By Vizing's theorem~\cite{BM76} for the ordinary edge coloring problem, 
we can color any bipartite graph with $\Delta$ colors.
For edge group coloring we can often do much better than this. 
Indeed, it's tempting to think that
we might be able to color any simple bipartite graph with $D$ colors.
Unfortunately, this is not true, as the example in Figure~\ref{example3} demonstrates.
In this figure, the shaded vertices are the outputs. 
Note that $D=2$, but this graph requires three colors.
We'll show in the next section that $D_i \Delta_o$ colors are always enough,
and we note that in \cite{KKP85}, the authors show that there are graphs that 
require more than $(D_i-1)\Delta_o$ colors.

The edge group coloring problem can be shown to be {\sl NP}-complete using a
reduction from the vertex coloring problem in graphs.
An instance of the vertex coloring problem is a graph $G=(V,E)$ and an integer $k$.
The objective is to determine if the vertices of $G$ can be colored with at most $k$
colors, with no two adjacent vertices having the same color.

\begin{figure}[h]
\centerline{\includegraphics[width=4.5in]{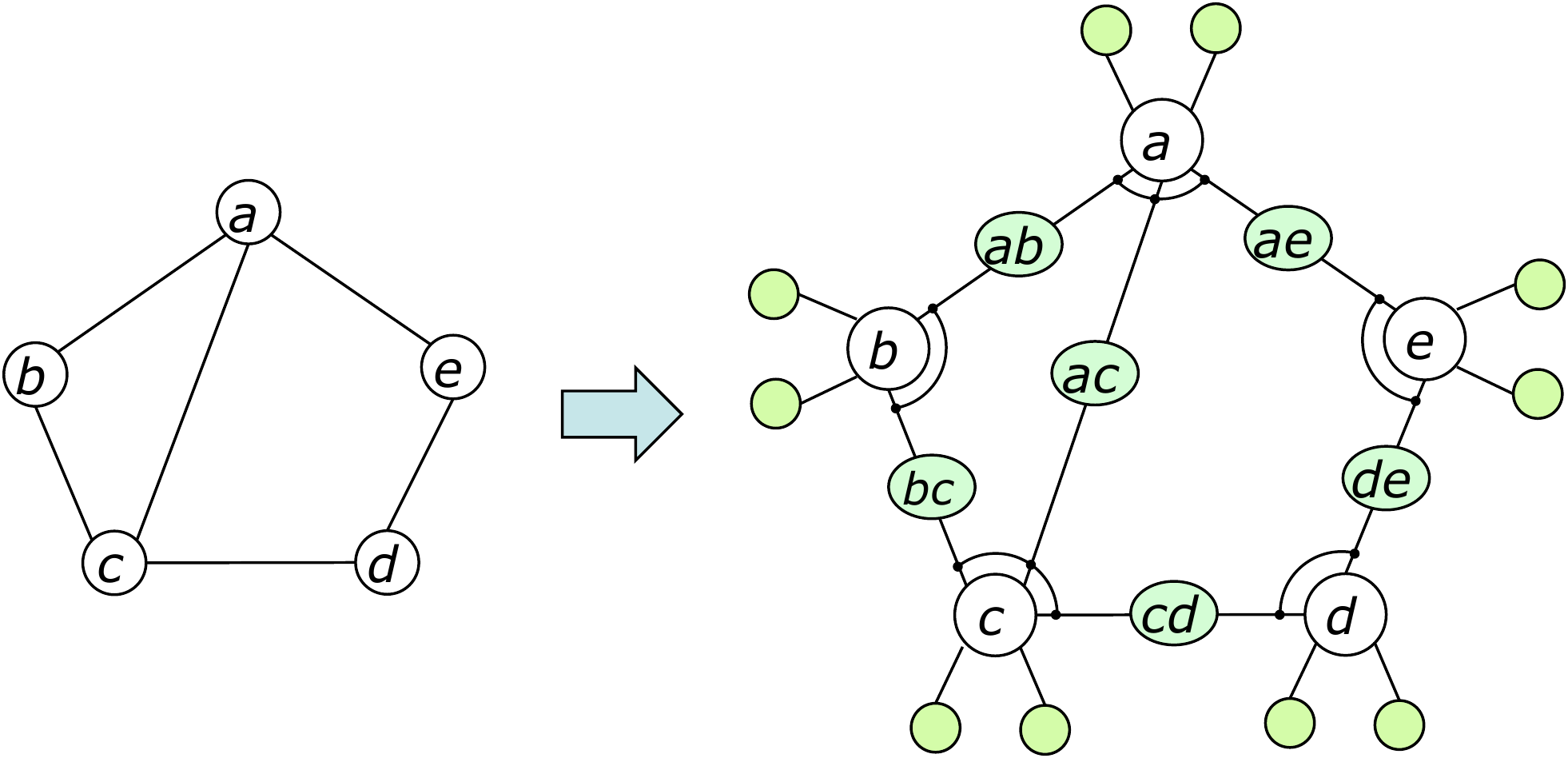}}
\caption{Reduction from graph coloring to edge group coloring}
\label{npReduction}
\end{figure}

Given an instance of the vertex coloring problem,
we can create a corresponding instance of the edge group coloring problem that
can be colored with $k$ colors if and only if the vertices of the original graph can
be colored with $k$ colors. To construct the edge group coloring instance, we start
with the original graph and insert a new vertex in the middle of each of its edges.
These new vertices are outputs in the edge group coloring instance, while
the original vertices are inputs. For each input $u$, all the edges incident to $u$
(so far) form a single edge group. To complete the construction, we add $k-1$ outputs
for each of the inputs. Each of these outputs is connected to its input by an edge.
We refer to these edges as {\sl stubs}. Each stub belongs to a singleton group.
An example of this reduction is
shown in Figure~\ref{npReduction}.

Note that in any valid
edge group coloring, the stubs consume $k-1$ colors, leaving one color for the
remaining group (corresponding to the edges in the original graph).
This means that there is a direct correspondence between proper $k$ colorings of
the original graph and edge-group colorings of the constructed graph.
This allows us to conclude that the edge group coloring problem is {\sl NP}-complete.
Moreover, this is true even for fixed values of $k$ as small as 3.

\section{A simple approximation algorithm}

We start with a simple observation. Consider a graph in which each
input has a single group. Such a graph can be colored with $\Delta_o$ colors,
since we can independently assign colors to the edges incident to each output.
This leads to the following general method for coloring the edge groups.
Repeat the following {\sl phase} until all edges have been colored.
\ipar
Select a set of edge groups (from among those not previously selected), with one group centered at each input
(omitting inputs whose edge groups have all been previously selected);
let $t$ be the maximum number of edges incident to
any single output in the subgraph induced by the
edges in the selected groups; color all the selected edges using $t$ previously unused colors.
\rapi
We call the groups selected in each phase a {\sl layer} and
the general method is called the {\sl layering method}.
The {\sl thickness} of a layer is the maximum number of edges incident to any output
in the subgraph defined by the layer.
Since the layer thickness cannot exceed $\Delta_o$, the number of colors used for a layer is
at most $\Delta_o$.
Since the number of phases is $D_i$, the number of colors used is at most $D_i \Delta_o$.
Moreover, since any valid coloring requires $\max \{D_i, \Delta_o\}$ colors,
this method produces solutions that use at most $\min \{D_i, \Delta_o\}$ times as many
colors as an optimal solution.

In the next section, we'll consider ways to refine the basic layering method,
but in the remainder of this section we evaluate its performance experimentally,
using a simple random graph model. The model uses five parameters:
the number of inputs, $n_i$; the number of outputs, $n_o$;
an upper bound on the group count at the inputs, $D_i$;
the vertex degree of the outputs, $\Delta_o$; 
and an upper bound on the number of colors needed to color the edges, $\chi$.
The generated graphs have input degree $\Delta_i=\Delta_o n_o/n_i$
(we require this quantity to be an integer). 
The color bound $\chi$ must be at least $\max\{D_i, \Delta_o\}$.
The generation process consists of three steps.
\begin{itemize}
\item Select a graph uniformly from among the bipartite graphs with the specified number of inputs and outputs, and the specified input and output degrees. 
\item At each output, randomly assign distinct colors in $\{1,\ldots,\chi\}$ to its incident edges.
\item At each input, form groups from the edges that share the same color, then for each input with more 
than $D_i$ groups, randomly merge groups until the number of groups is $D_i$.
\end{itemize}
The coloring used to generate the group graph is discarded once the generation process is complete.
The minimum number of colors required for these graphs falls in the range $[\max\{D_i, \Delta_o\}, \chi]$.
An example of a group graph generated using this method is shown below.
\ipar
\begin{verbatim}
[a: (f i l) (g k) (e)]
[b: (i l) (h j) (g k)]
[c: (f h j) (e) (g h)]
[d: (f i) (e j) (k l)]
\end{verbatim}
\rapi
Each line in this representation shows the neighbors of an input, with the parentheses
identifying the groups. So for example, input $a$ has three groups, the first consisting
of the edges $(a,f)$, $(a,i)$ and $(a,l)$. This graph was constructed using $\chi=4$
and a 4-coloring appears below.
\ipar
\begin{verbatim}
1: a(f i.) b(h j) c(e) d(k l)
2: a(g k) b(i l) c(h j.) d(f.)
3: a(l.) b(g.) c(f.) d(e j)
4: a(e) b(k.) c(g h) d(i.)
\end{verbatim}
\rapi

Here, each line identifies the edges assigned a particular color, with parentheses again
used to identify groups or partial groups (which are indicated using a period).
So for example, the first two edges in $a$'s first group are assigned color 1, while the
remaining edge is assigned color 3.
The coloring obtained by the basic layering method for this graph uses seven colors and is shown below.
\ipar
\begin{verbatim}
1: a(f i l) c(h j.)
2: b(i l) c(f.)
3: d(f i)
4: a(g k) b(h j) c(e)
5: d(e j)
6: a(e) b(g k) c(h.) d(l.)
7: c(g.) d(k.)\end{verbatim}
\rapi
The first layer selected by the algorithm includes the first group at each input and has
a thickness of 3. Hence, the first three colors are used to color its edges. The next layer
is colored with the next two colors and the third layer with the last two.
\begin{figure}[t]
\centerline{\includegraphics[width=4in]{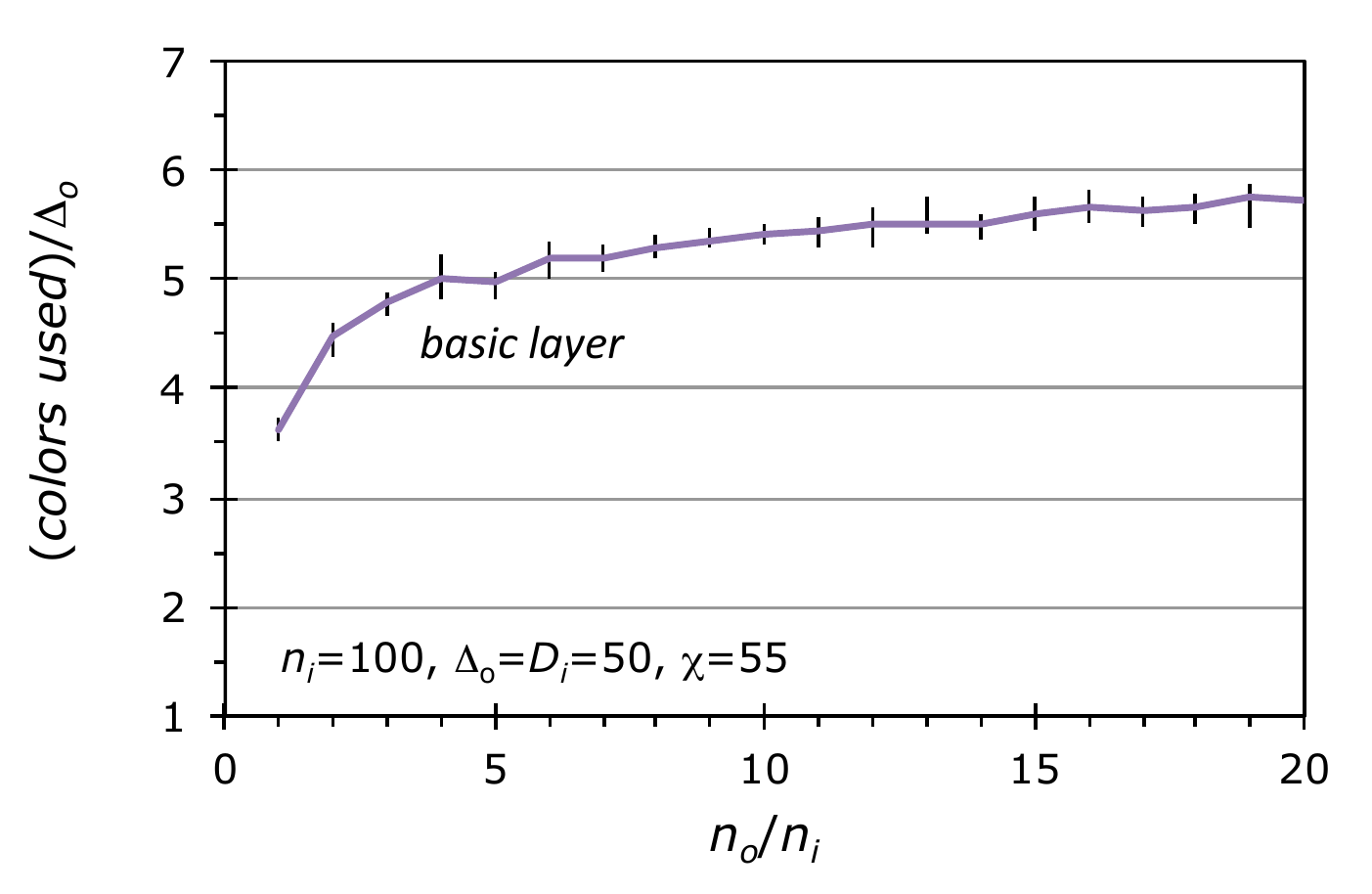}}
\caption{Effect of input/output asymmetry on basic layering method}
\label{basicLayersAsymmetry}
\end{figure}

Our first experiment explores the effect of input/output {\sl asymmetry} 
($n_o/n_i$) on the performance
of the basic layering method. For this experiment, we use dense graphs, where the number of edges $m=n_i n_o/2$.
The bound on the number of groups per input is equal to the output degree
($D_i=\Delta_o$) and
the color bound is slightly larger than the output degree ($\chi = 1.1\Delta_o$).
The chart in Figure~\ref{basicLayersAsymmetry} shows the ratio of the number of colors used to $\Delta_o$.

The curve shows the average of ten trials on different random group graphs. 
The error bars show the minimum and maximum values from these ten trials. 
We observe that the number of colors increases initially as the graph becomes
asymmetric, but then the rate of increase tapers off. 
For the most asymmetric graphs, the number of colors used is just
under six times the lower bound, $\Delta_o$.
(The upper bound on the number of colors is $50\Delta_o$ in this case.)

\begin{figure}[t]
\centerline{\includegraphics[width=4in]{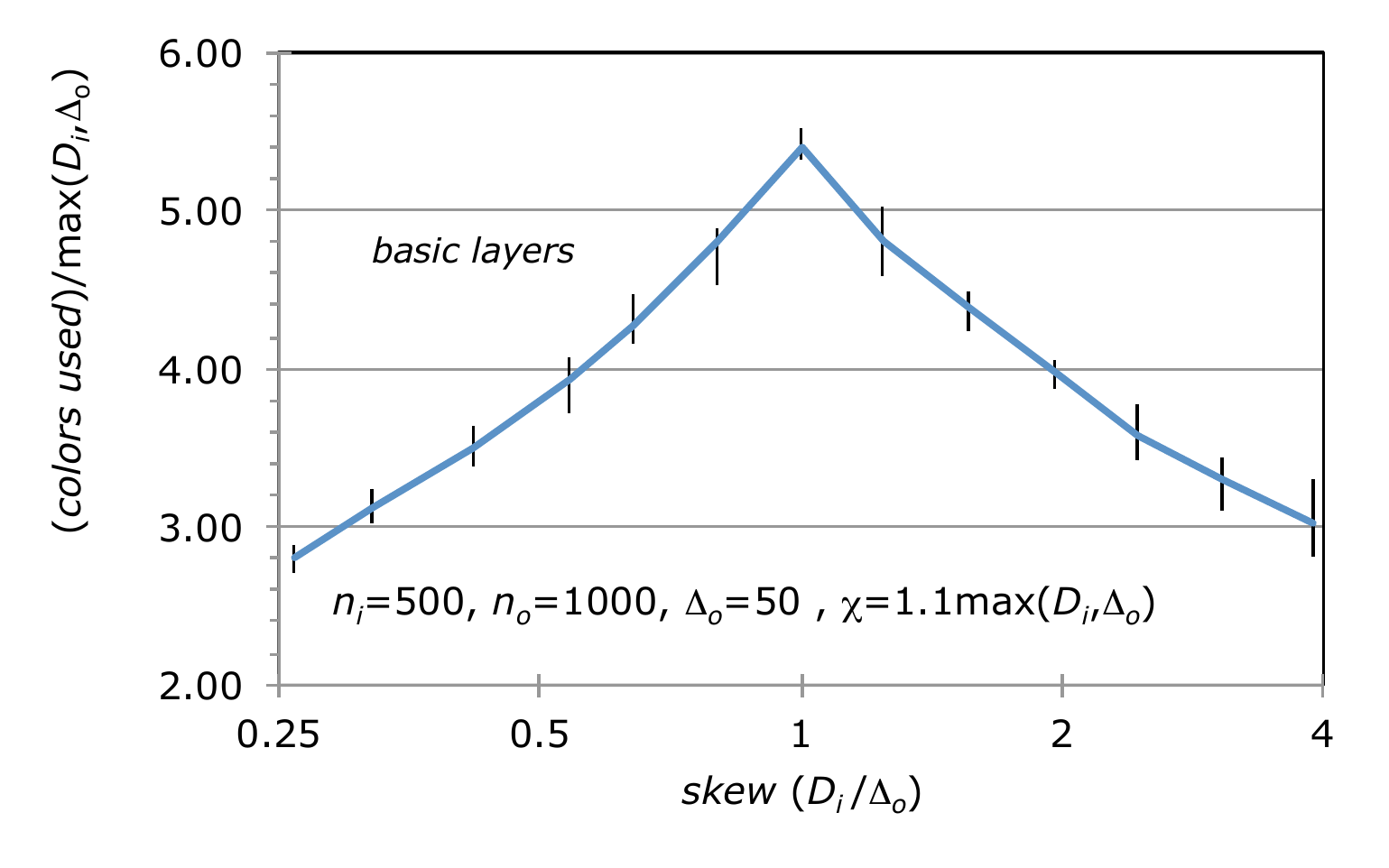}}
\caption{Effect of skew on basic layering method}
\label{basicSkew}
\end{figure}

Our next experiment examines how the performance varies as a function of
the ratio of $D_i$ to $\Delta_o$, which we refer to as the {\sl skew}.
The results are shown in Figure~\ref{basicSkew}.
Observe that the performance ratio is highest (worst) when $D_i=\Delta_o$.
In general, we can expect better performance (relative to the lower bound)
when $D_i$ and $\Delta_o$ differ substantially.

\begin{figure}[t]
\centerline{\includegraphics[width=4in]{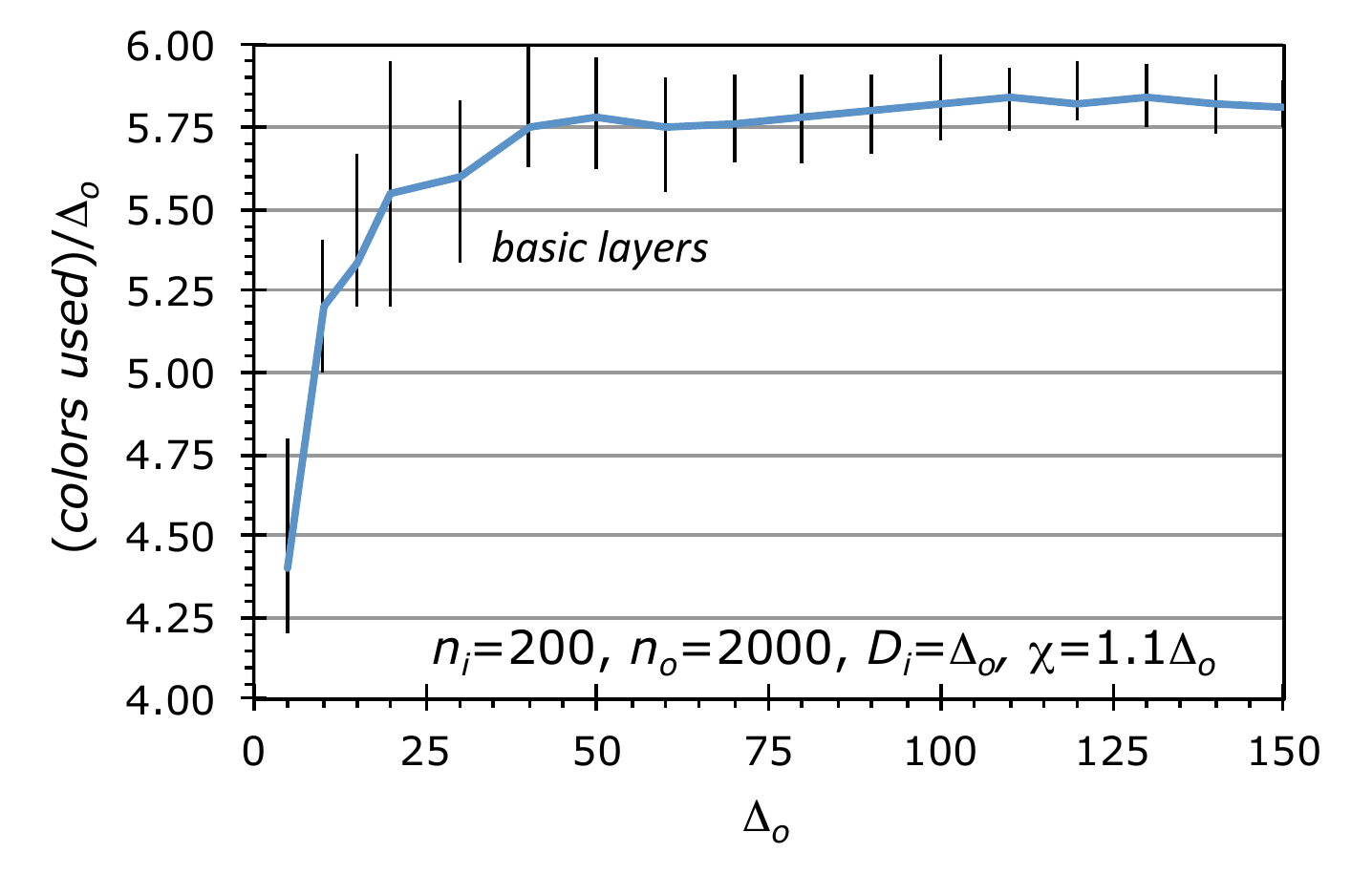}}
\caption{Effect of graph density on basic layering method}
\label{basicLayersDensity}
\end{figure}

In our next experiment, we examine how the performance varies with
the density of the graph.
In Figure~\ref{basicLayersDensity} we hold the number of vertices fixed, while increasing the
degree of the inputs and outputs. 
Observe that there is a modest increase in the performance ratio as $\Delta_o$ increases from 5 to 40,
but after that, it remains roughly constant. Since the number of phases of the algorithm increases directly
with the graph density, the performance ratio is just a reflection of the layer thickness, which is
only weakly dependent on density.

\begin{figure}[t]
\centerline{\includegraphics[width=4in]{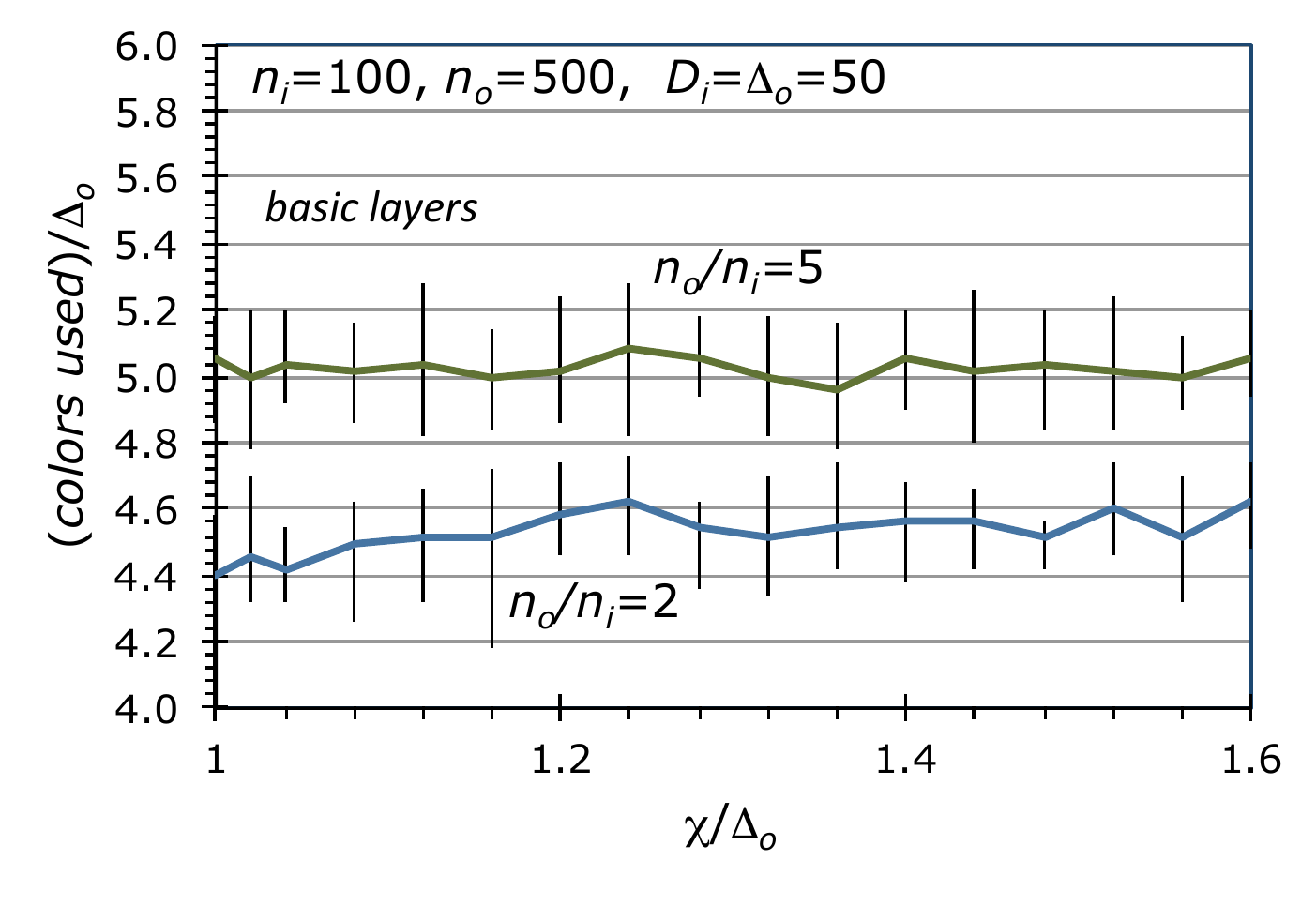}}
\caption{Effect of color bound on basic layering method}
\label{basicLayersBound}
\end{figure}

Our final experiment for the {\sl basic layer} method examines the impact of the color bound $\chi$. One might expect that increasing $\chi$ would also increase the number of colors
required, and the number used by the algorithm.
Figure~\ref{basicLayersBound} shows that, there is
no clear relationship between $\chi$ and the number of colors used by the algorithm.
On reflection, this is not surprising.
Recall that the number of iterations is $D_i$ and the number
of colors used is $D_i$ times the average layer thickness. Neither $D_i$, nor the layer thickness is 
directly affected by the choice of $\chi$, so neither is the performance of the algorithm.
(Note that the error bars in this chart are exaggerated by the limited scale of the $y$-axis.)

The overall conclusion is that the performance of the basic layering method is primarily a function of the
layer thickness, which is most strongly influenced by input/output asymmetry, although even in that case the
dependence tapers off as asymmetry grows beyond 5.
We also observe that group graphs with $D_i=\Delta_o$ require the largest
number of colors.

\section{Refining the layering method}

In this section we introduce several refinements to the basic layering method and observe how they
affect its performance.
Our first refinement is called the {\sl thin layers} method, since it seeks to select the groups that
form each layer with the objective of minimizing the overall layer thickness. More precisely, the
{\sl thin layers} method forms each layer by repeating the following step
at each input $u$.
\ipar
Select the group at $u$ that yields the smallest thickness value when added to the layer 
formed by the groups that have been selected so far.
\rapi
Once the groups forming a layer have been selected, they are colored using $t$ new colors, where
$t$ is the thickness of the new layer.
When applied to the example graph from the last section, the thin layers method produces the
coloring using six colors shown below.
\ipar
\begin{verbatim}
1: a(l f i) b(j h) c(e)
2: d(i f)
3: a(k g) b(l i) c(j f h)
4: d(j e)
5: a(e) b(k g) c(h.) d(l.)
6: c(g.) d(k.) \end{verbatim}
\rapi
Here, each of the three layers has a thickness of 2.

Before describing our next algorithm, we need a definition. 
Given a partial coloring of a graph, a color $c$ and an uncolored edge $e=(u,v)$, 
we say that $c$ is {\sl viable for $e$ at $u$} if $c$ is not being
used by any other edge at $u$ that is in a different group than $e$;
$c$ is viable for $e$ at $v$ if it is not being used by any other edge at $v$.
We say that $c$ is {\sl viable for $e$} if it is viable at both endpoints.

Our next method extends the thin layers method by relaxing the requirement that each layer use a distinct set of colors.
The {\sl min color} method colors each selected
group using colors allocated for previous layers whenever possible.
In particular, when coloring a group, each of its edges $e$ is colored
according to the first of the three cases listed below, that is applicable.
\begin{enumerate}
\item
If there is a color $c$ that is viable for $e$ and has already been used by some edge in $e$'s group,
color $e$ using the smallest such color (that is, the color with the smallest index).
\item
If there is a color $c$ that is viable for $e$ and has already been used by any edge,
color $e$ using the smallest such color.
\item
If there is no previously used color that is viable for $e$, allocate the next unused color
and use it to color $e$.
\end{enumerate}
Colors are allocated in increasing order of their positive integer index.
When applied to the example graph from the last section, the {\sl min color} method produces the
coloring using five colors shown below.
\ipar
\begin{verbatim}
1: a(f i l) b(h j) c(e) d(k.)
2: a(g k) b(l.) c(h j.) d(f i)
3: b(i.) c(f.) d(e j)
4: a(e) b(g k) c(h.) d(l.)
5: c(g.) \end{verbatim}
\rapi
Here, the first and second layers each require two new colors, 
but the third layer is colored using just one additional color.

Our last refinment to the basic layers method is inspired by the classical proof of Vizing's theorem
for the standard edge coloring problem. 
The proof describes an algorithm that uses {\sl augmenting paths} to
find a coloring. 
That method cannot be used directly for the edge group coloring problem, but we can adapt it
to accommodate edge groups.

\begin{figure}[ht]
\centerline{\includegraphics[width=4in]{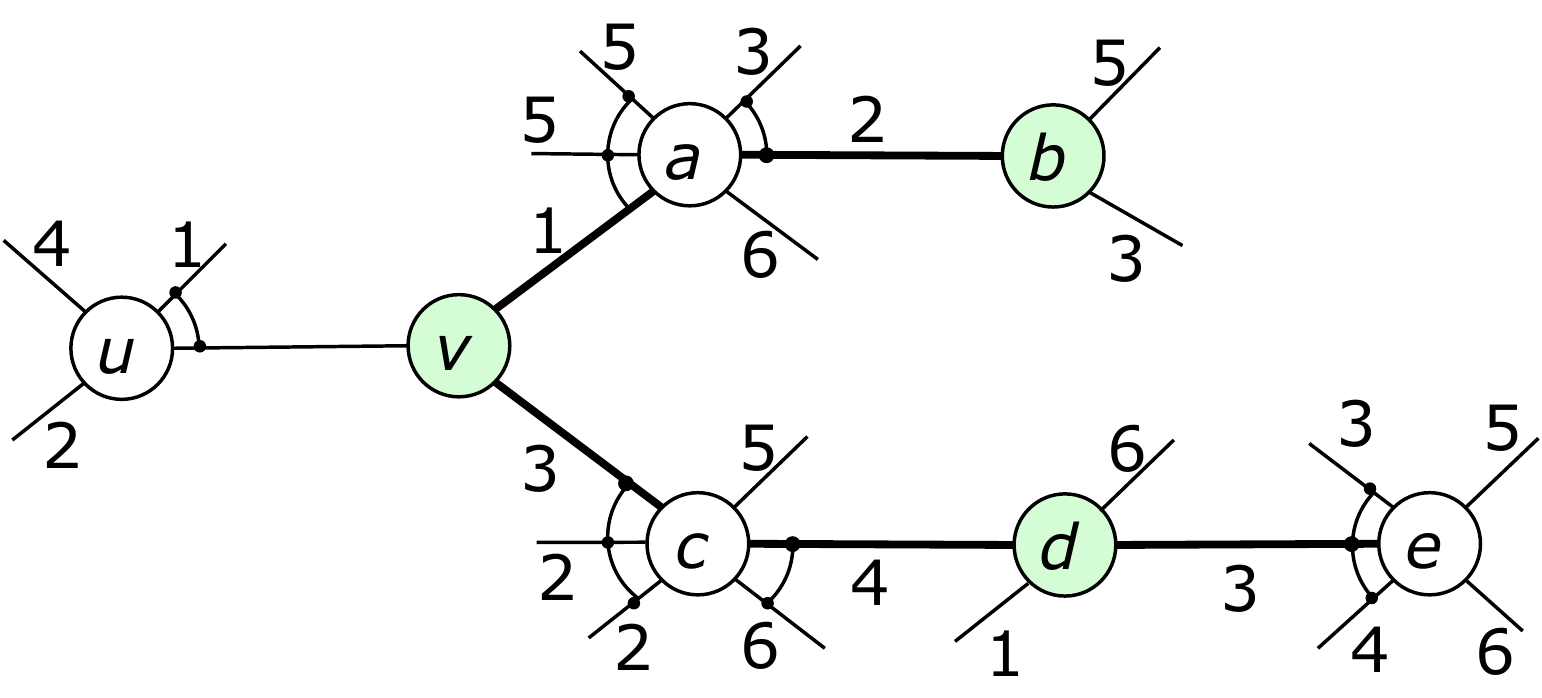}}
\caption{Example of augmenting paths}
\label{augPath}
\end{figure}

Let $e=(u,v)$ be an edge to be colored and let
$i$ and $j$ be two edge colors where $i$ is viable for $e$ at $u$,
$j$ is viable at $v$, but neither is viable at both endpoints.
An {\sl augmenting path} is a path $p$ that starts at $v$, ends at a vertex $w$, 
has edges that alternate in color between $i$ and $j$ (starting with $i$) and that satisfies the
following conditions.
\begin{itemize}
\item
The two path edges incident to any intermediate vertex $x$ belong to
different groups.
\item
No non-path edge incident to an intermediate vertex $x$ has color $i$ or $j$.
\item
If colors $i$ and $j$ are both in use at $w$, the edges with those colors
are all in the same group.
\end{itemize}
Observe that if $p$ is an
augmenting path, we can reverse the colors of the edges on $p$ and
still have a valid coloring. Performing this color reversal makes color $i$
viable for $e$. Figure~\ref{augPath} shows an example of a graph with 
two augmenting paths, a 1-2 augmenting path from $v$ to $b$ and a 3-4-augmenting
path from $v$ to $e$.

Note that when attempting to construct an augmenting path, we may arrive at a
vertex that does not satisfy either the conditions for an intermediate vertex $x$, or
the conditions for the terminal vertex $w$.
In this case, there is no $ij$-augmenting path for $e$
and the attempted path construction fails.

The {\sl recolor} method is a modification of the {\sl min color} method. In particular,
when coloring edge $e=(u,v)$, it replaces the third case in the {\sl min color} method with the following.
\ipar
If there is no previously used color that is viable for $e$, color $e$ using
the first applicable sub-case listed below (this may involve testing multiple color pairs).
\begin{itemize}
\item
If color $i$ is used by some other edge in $e$'s edge group at $u$, color $j$ is a previously used color
that is unused at $v$ and there is an $ij$-augmenting path from $v$, 
reverse the colors of the edges on the path and and color $e$ using $i$.
\item
If color $i$ is a previously used color that is unused at $u$, color $j$ is a previously used color
that is unused at $v$ and
there is an $ij$-augmenting path from $v$, reverse the colors of the edges on the path and
color $e$ using $i$.
\item
If neither of the previous sub-cases apply for any pair of previously used
colors, allocate the next unused color and use it to color $e$.
\end{itemize}
\rapi
Observe that in the example in Figure~\ref{augPath}, the {\sl recolor} algorithm will choose
the 1-2 augmenting path in preference to the 3-4 path.

\begin{figure}[t]
\centerline{\includegraphics[width=4in]{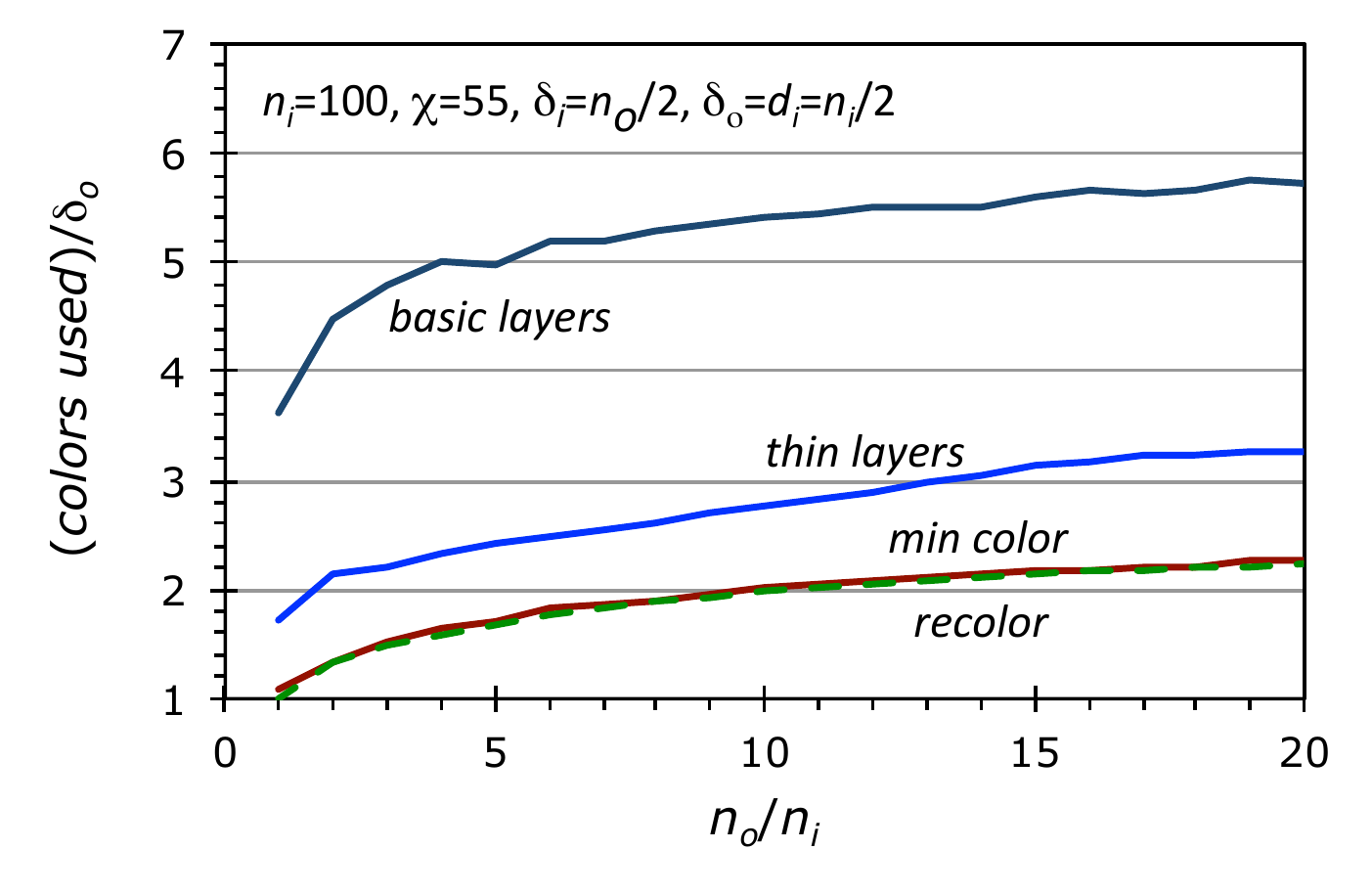}}
\caption{Effect on input/output asymmetry on the layering algorithms}
\label{layersAsymmetry}
\end{figure}

Next, we examine the performance of our refinements to the basic layering method.
Figure~\ref{layersAsymmetry} shows how the performance varies with asymmetry
(we have omitted the error bars for clarity).
Observe that for the most asymmetric graphs, the {\sl thin layers} method brings the performance ratio
down from about 5.8 to about 3.3. The {\sl min color} and {\sl recolor} methods 
bring it down
to about 2.2. The {\sl recolor} method provides only a very small improvement over the {\sl min color} method
(a dashed line style is used for {\sl recolor} to make the small difference more apparent).
Finally note that for asymmetries smaller than 10, the min color and recoloring methods produce solutions
that are within a factor of two of optimal.

\begin{figure}[t]
\centerline{\includegraphics[width=4in]{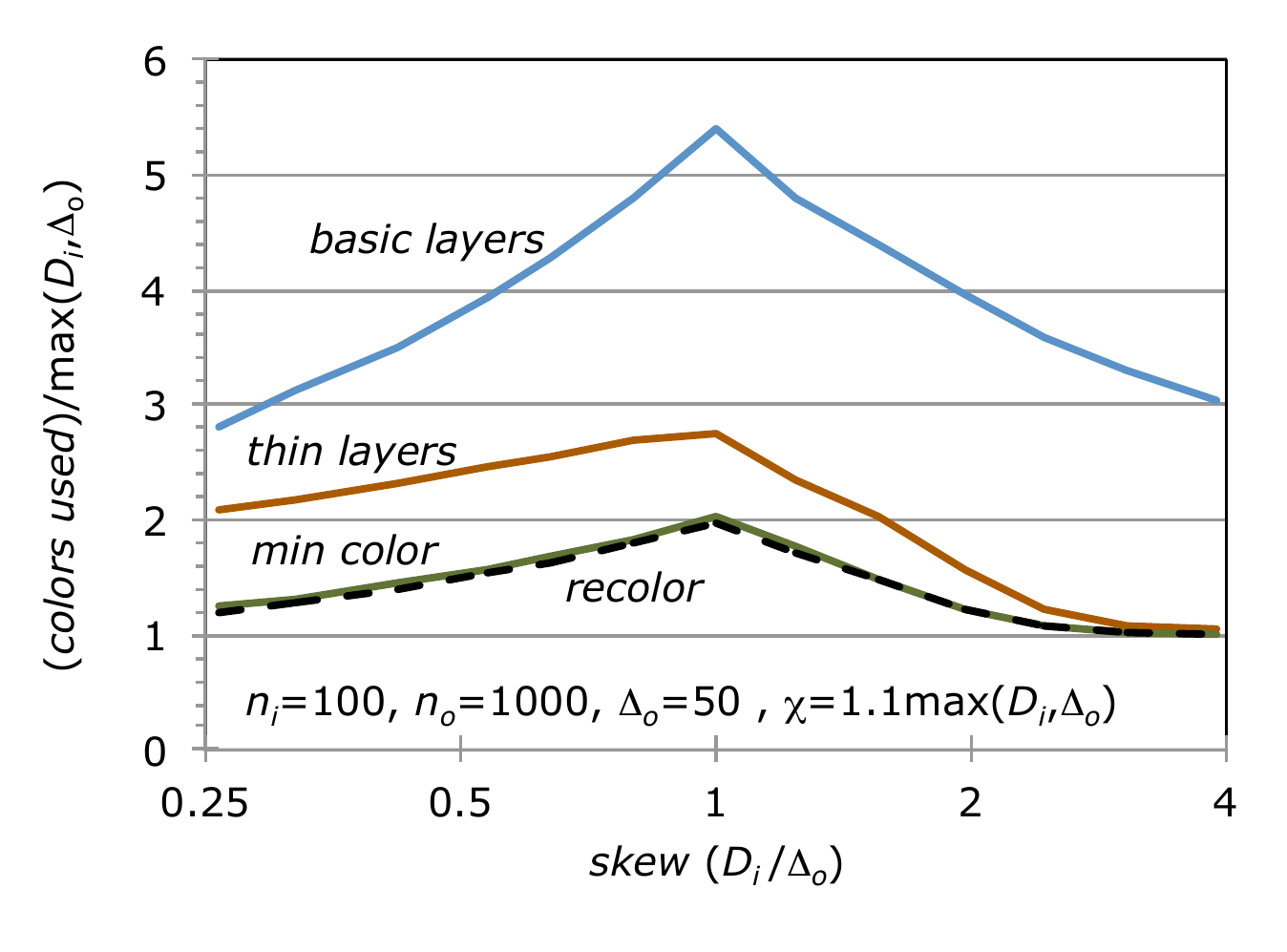}}
\caption{Effect of skew on the layering algorithms}
\label{layersSkew}
\end{figure}

\begin{figure}[t]
\centerline{\includegraphics[width=4in]{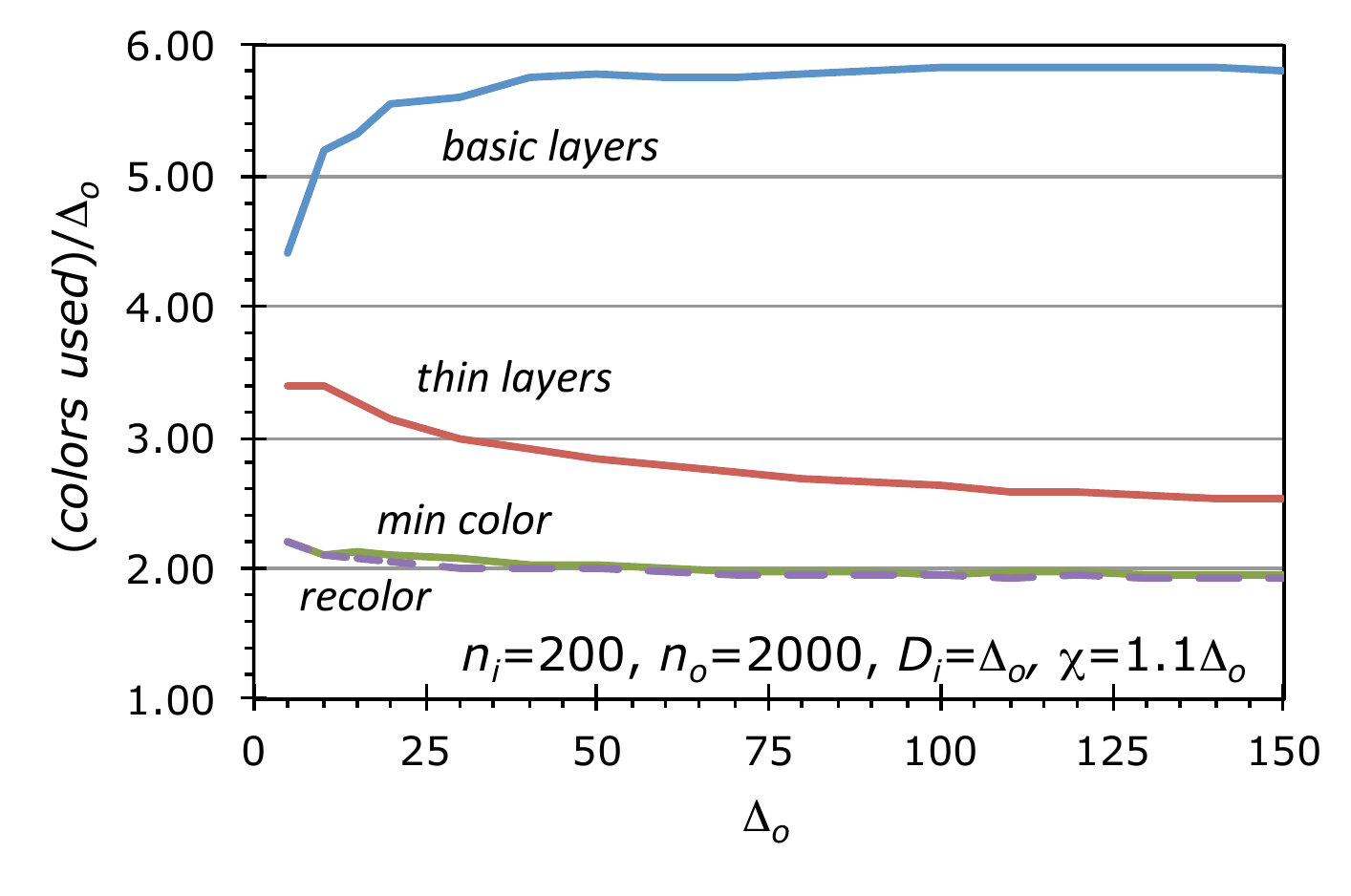}}
\caption{Effect of graph density on the layering algorithms}
\label{layersDensity}
\end{figure}

Figure~\ref{layersSkew} shows how the performance varies with skew.
In all cases, we see a peak when $D_i=\Delta_o$, although the height of the
peaks is smaller for the more complex variations.
Figure~\ref{layersDensity} shows how the performance ratio is affected by the graph density.
Here, we observe that for the algorithms introduced in this section, the average number of colors used per layer decreases (by small amounts) as the density increases. 
There appear to be two factors accounting for this.
First, as density increases, we have more groups per input, which provides a wider range of available choices when forming the layers.
This leads to thinner layers, particularly in the early phases.
The second factor that plays a role arises only in the {\sl min color} and {\sl recolor}
methods. Because dense graphs require more colors, they offer a wider range of color choices that the algorithms can exploit when coloring the individual edges.

\section{Re-visiting an algorithm by Yang and Masson}

In ~\cite{YM91}, Yang and Masson describe an algorithm for routing a single multicast connection
in a three stage Clos network. They show that a suitably configured Clos network is wide-sense
nonblocking for new multicast connections, if their algorithm is used.
Here, we adapt their algorithm for the edge-group coloring problem.
The algorithm has a parameter $k$ that limits the number of colors used for any single
edge group (we'll discuss the choice of $k$ later).
We say that a color is {\sl eligible} for
selection at some stage, if its index is $\leq \max\{D_i,\Delta_o\}$ or it has already been used at least once (possibly in an earlier step).
The algorithm attempts to color each edge group with at most $k$ colors,
by selecting colors using a greedy strategy. 
\ipar
While some edge in the group remains uncolored, select an eligible color $c$ that is viable 
for the largest number of uncolored edges remaining in the group;
use $c$ to color all edges in the group for which it is viable.
\rapi
If this procedure fails to color all edges in a group using $\leq k$ colors, we allocate a new color and use it to color all edges of the group, instead.

We refer to this as the {\sl few colors} method.
When applied to the graph
\ipar
\begin{verbatim}
[a: (f i l) (g k) (e)]
[b: (i l) (h j) (g k)]
[c: (f h j) (e) (g h)]
[d: (f i) (e j) (k l)]\end{verbatim}
\rapi
the few colors method produces the 5-coloring shown below, when $k=2$, assuming that
the edge groups are colored in the order in which they appear in the graph.
\ipar
\begin{verbatim}
1: a(f i l) b(h j) c(e) d(k.)
2: a(g k) b(i l) c(f h j)
3: a(e) b(g k) d(f i)
4: c(g h) d(e j)
5: d(l.)\end{verbatim}
\rapi

The number of colors used by the few colors method is bounded in the following theorem,
which is an adaptation of a theorem in~\cite{YM91}.

\begin{theorem}
Let $G=(V,E)$ be a bipartite graph in which the edges incident to each input are
divided into edge groups to produce an instance of the edge group coloring problem.
The few colors algorithm with parameter $k$ colors the edges of $G$ using at most
$$
(D_i -1)k + (\Delta_o -1) n_o^{1/k} +1
$$
colors.
\label{fewColorsTheorem}
\end{theorem}
We give a simplified proof of the theorem, based on a lemma that bounds the number of colors used by
a greedy algorithm for the set covering problem, when applied to instances that satisfy a special condition.
An instance of the set covering problem consists of a base set $A$ and a collection of subsets $S_i$ of $A$.
The objective is to find a smallest possible collection of subsets whose union includes all elements of $A$.
The greedy algorithm for the set covering problem repeatedly selects a subset that covers the
largest possible number of previously uncovered elements. It terminates as soon as all elements
are included in at least one of the selected subsets.
\begin{lemma}
Let $A=\{a_1,\ldots,a_t\}$ and $S=\{S_1,\ldots,S_p\}$
 be an instance of the set covering problem in which every $a_i$  
appears in at least  $p-q$ subsets for some  integer $q$. 
If integer $k$ satisfies $p>qt^{1/k}$ then the greedy algorithm finds
a covering of $A$ using at most $k$ subsets.
\end{lemma}

\noindent
{\sl proof}. 
Let $h$ be the number of subsets in the greedy solution and assume the subsets are numbered so that 
for $1\leq i \leq h$, $S_i$ is the subset chosen in step $i$.
Define $U_i=S_1 \cup \ldots \cup S_i$  and let $D_i=S_i - U_{i -1}$, 
$s_i=|S_i|$, $u_i =|U_i|$, $d_i= |D_i|$.
Then,
$$
ps_1 \geq\sum_{i=1}^p s_i \geq (p-q)t
$$
So, $u_1=s_1\geq(1-q/p)t=(1-x_o)t$,
where $x_j=(q-j)/(p-j)$. Next, note that
$$
(p-1)d_2 \geq\sum_{i=2}^p |S_i-U_1|\geq(p-q)(t-u_1)
$$
So, $d_2\geq(1-x_1)(t-u_1)$ and
$$
u_2=u_1+d_2\geq (1-x_1)t +x_1u_1 \geq (1-x_1)t+x_1(1-x_0)t=(1-x_1x_0)t
$$ 
Extending this reasoning using induction, 
we find that for $i\leq h$, $u_i\geq (1-x_{i-1} \cdots x_1 x_0)t$. In particular
$$
u_k\geq (1-x_{k-1} \cdots x_1 x_0)t \geq (1-x_0^k)t>(1-1/t)t = t-1
$$
So, $U_k$ has more than $t-1$ elements and since $|A|=t$, $|U_k|=t$. $\Box$

Now, let's proceed to the proof of the theorem.
We view each step of the {\sl few colors} method as a set covering problem, 
in which the base set is the set of edges in the current edge group.
There is a subset for each color that is not already being used by 
an edge incident to the group's input $u$, and for each such color $c$, 
its subset consists of those edges $(u,v)$ for which $c$ is not being used by any edge incident to $v$.
Let $t$ be the number of edges in the group and let $p$ be the
number of subsets.

Now, let $C$ be the number of colors that have been used so far.
We claim that if $C>(D_i -1)k + (\Delta_o -1)n_o^{1/k}$, the group can
be colored with $\leq k$ colors, without any further increase in $C$. 
So, assume $C>(D_i -1)k + (\Delta_o -1)n_o^{1/k}$ at the start of the step.
Let $u$ be the group's input and note that the number of colors already in use at $u$ is $\leq(D_i-1)k$.
Hence, the number of colors that are available to color the group is 
$\geq C-(D_i -1)k>(\Delta_o -1)n_o^{1/k}$. That is $p>(\Delta_o -1)n_o^{1/k}$.

Next, consider any edge $(u,v)$ in the group. At most $\Delta_o-1$ colors are already being
used at $v$. Consequently, $(u,v)$ is a member of at least $p-(\Delta_o -1)$ subsets.
Letting $q=(\Delta_o-1)$, we have $p>qn_o^{1/k}$ and since $t\leq n_o$, $p>qt^{1/k}$.
The lemma then implies that the group can be colored using at most $k$ colors. $\Box$

We may choose the value of $k$ that minimizes the bound in the theorem.
When $D_i=\Delta_o$, the best choice of $k$ is the smallest integer for which
$$
 n_o^{1/k} - n_o^{1/k+1} \leq 1
$$
The required value is approximately $2\ln n_o/\ln\ln n_o$, as shown by Yang and Masson.
This also leads to a worst-case performance ratio of approximately $(2\ln n_o/\ln\ln n_o) + (\ln n_o)^{1/2}$.
For all but the sparsest graphs, this is better than the trivial ratio for the layering method.

Note that the performance ratio is independent of the order in which the edge groups 
are selected. We find that coloring the groups in decreasing order of their size
gives better performance than other simple ordering strategies.
This approach allows the largest groups to be colored with fewer colors, leaving
more colors available for groups that come later.
We can also refine the method in another way. Instead of using a fixed parameter $k$,
we can allow the number of colors per group to vary with the number of groups at 
an input.
Specifically, we allow a group centered at an input $u$ to use up to
$$\lceil \hbox{(the number of eligible colors at $u$)}/d(u) \rceil$$
colors. Recall that the number of eligible colors is initially $\max\{D_i,\Delta_o\}$
and it increases by one whenever a previously unused color is used.
We evaluate this version of the {\sl few colors} method experimentally at
the end of the next section.

\section{Using matchings to color group graphs}

In~\cite{KKP85}, the authors give several results that can be interpreted as bounds on
the number of colors needed to color a group graph. 
Their main result is that the number of colors required for any graph is at most
$$(D_i -1)\lfloor \log_2 (2n_o)\rfloor + 2\Delta_o$$
For most parameter choices, this is not quite as strong as Yang and Masson's
result, but the method used to prove it leads to an interesting and
distinctly different algorithmic method.
In this section, we describe this general method and evaluate two specific
algorithms based on it.
The method divides the coloring process into two parts. First,
at each input, we divide the available colors among the groups at that input.
The colors assigned to a given group are referred to as its {\sl menu}.
Next, we color the edges incident to each output using colors selected
from the menus of the groups that the edges belong to. We can illustrate this
procedure for the example graph
\ipar
\begin{verbatim}
[a: (f i l) (g k) (e)]
[b: (i l) (h j) (g k)]
[c: (f h j) (e) (g h)]
[d: (f i) (e j) (k l)]
\end{verbatim}
\rapi
Suppose we are attempting to color this using four colors.
In the first step, we create menus for each group.
\ipar
\begin{verbatim}
[a: (f i l)[1,2] (g k)[3] (e)[4]]
[b: (i l)[1] (h j)[2] (g k)[3,4]]
[c: (f h j)[1,2] (e)[3] (g h)[4]]
[d: (f i)[4] (e j)[2,3] (k l)[1]]
\end{verbatim}
\rapi
The menus are shown within the square brackets, so for example
the menu of the group {\tt a(f i l)} contains colors 1 and 2.
In the next step, we attempt to color the edges at each output using the
colors in the menus associated with its edge groups.
For example, output $e$ has three edges $(a,e)$, $(c,e)$ and $(d,e)$
whose groups have menus $[4]$, $[3]$ and $[2,3]$;
so, they can be colored using colors 4, 3 and 2.
Similarly, the edges incident to output $f$ can be colored using colors 1, 2 and 4.
When we get to output $g$, we find that the groups for its three edges have
menus $[3]$, $[3,4]$ and $[4]$. Since these menus give us only two colors to
choose from, we cannot color all three edges, given these menus.
This raises the question of how we can best select the menus in the first place.
We'll explore two different ways to answer this question and the algorithms
based on those answers. First however, let's consider how to color the edges
once the menus have been chosen. This involves finding a matching in a graph.

\begin{figure}[ht]
\centerline{\includegraphics[width=1.5in]{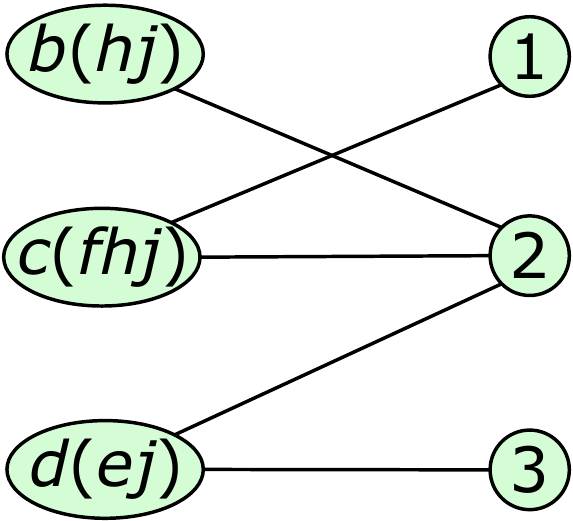}}
\caption{Menu graph for output $j$}
\label{menuGraf}
\end{figure}

For each output $v$, define $M(v)$ to be the {\sl menu graph} of $v$.
$M(v)$ has an input for each group with an edge incident to $v$ and
an output for every color.
It includes edges joining each input to the outputs corresponding to colors in the 
group's menu.
So in the earlier example, $M(j)$ has the menu graph in Figure~\ref{menuGraf}.
A matching that includes an edge incident to every input of $M(v)$
defines a valid coloring of the edges incident to $v$.
We call such a matching {\sl complete}.
So given an assignment of colors to menus, we can color all the edges of
the graph if we can find complete matchings for all the menu graphs.
If any menu graph does not have a complete matching, then there is
no valid coloring using the given menus.

We start by considering random menus. Here, we assign $k$ colors
to each group, for some integer $k$. We can show that for large enough values of $k$,
this approach succeeds with high probability. By Hall's Theorem~\cite{BM76},
a random menu graph with $k$ edges incident to each input has a complete matching
if and only if, for every non-empty set $X$ of the inputs, 
the set of neighbors of $X$
is at least as large as $X$. For large enough $k$,
the probability that a random menu graph contains a set of inputs
$X$ for which the set of neighbors is smaller than $X$ is small. 
The required analysis is a variation on well-known
results for expander graphs~\cite{HLW06}.

\begin{theorem}
Let $G$ be a group graph and let $C=k \max(D_i,\Delta_o)$, where
$k\geq 2\left(\ln 2\Delta_o n_0/\ln\ln 2\Delta_o n_0\right)^{1/2}$.
A set of random menus using $C$ colors and $k$ colors per group
defines a set of menu graphs that all contain a complete matching 
with probability $\geq 1/2$.
\end{theorem}

{\sl proof.}
Consider the menu graph $M(v)$ for some vertex $v$.
Let $X$ be a subset of the inputs of $M(v)$ and $Y$ be a subset of its outputs.
Let $N_{X,Y}$ be the event that all of $X$'s neighbors are in $Y$. Then
$$\Pr \{N_{X,Y}\} = \left(|Y|/C\right)^{k|X|}$$
The probability that there exists a pair of such subsets $X$, $Y$ where $|Y|<|X|$ is
\begin{eqnarray*}
\leq \sum_{i={k+1}}^{\delta(v)} {\delta(v) \choose i}{C \choose i-1} 
		\left(\frac{i-1}{C}\right)^{ik} 
&\leq& \sum_{i={k+1}}^{\Delta_o} {\Delta_o \choose i}{C \choose i} 
		\left(\frac{i}{C}\right)^{ik} \\
&\leq& \sum_{i={k+1}}^{\Delta_o} \left(\frac{e \Delta_o}{i}\frac{eC}{i}
		\frac{i^k}{C^k}\right)^{i} \\
&\leq& \sum_{i=k+1}^{\Delta_o} \left(\frac{e^2\Delta_o i^{k-2}}{C^{k-1}}\right)^{i} \\
&\leq& \Delta_o \left(e^2\left(\frac{\Delta_o}{C}\right)^{k-1} \right)^{k+1} \\
&\leq& \Delta_o \left(e^2/k^{k-1} \right)^{k+1}
\end{eqnarray*}
The right side of this last expression is $\leq 1/2n_o$ if $k$ satisfies the 
condition in the theorem.
Since the number of menu graphs is $n_o$, the probability that the {\sl random menu}
method fails to find full matchings for all menu graphs is $\leq 1/2$. $\Box$

The {\sl random menu} method selects random menus repeatedly until it finds
a set for which the menu graphs all have complete matchings.
If $k$ satisfies the bound in the theorem, we can expect to succeed after a
small number of attempts. In our experimental implementation of the
{\sl random menu} method, we allow inputs with $<D_i$ groups to have
more colors per group.
In particular, at each input, we assign colors to groups in a round-robin fashion, 
starting with the largest group (consequently, larger groups may get one more color than the smallest groups).
For the most asymmetric graphs used in our experiments, the bound in the
theorem requires $k\geq 5$.
We find that for random graphs, 2 or 3 colors per group is generally sufficient.

We can also construct menus in a more systematic way. For each output, we maintain
a menu graph and a maximum matching on that graph. We define the {\sl deficit}
of a group $g$, to be the number of menu graphs in which there is a vertex for $g$
which is not currently matched. We define the {\sl gain} of a color $c$ not in $g$'s menu, to be the number of menu graphs  in which the vertices for $g$ and $c$ are both
unmatched. The gain is a lower bound
on the reduction in $g$'s deficit that will result if $c$ is added to its menu.

The {\sl greedy menu} method
repeatedly performs the following step, until all groups have a zero deficit.
\ipar
Select a group $g$ with a positive deficit. Let $u$ be the input on which $g$ is centered
and let $k_g=\lceil (\hbox{number of eligible colors})/ d(u)\rceil$.
While $g$ has a positive deficit, its menu has fewer than $k_g$ colors,
and some previously used color has positive gain for $g$, select one such color and 
add it to $g$'s menu; update all menu graphs containing a vertex for $g$ and
update their maximum matchings.
If $g$ still has a positive deficit, allocate a previously unused color, remove all colors
currently in $g$'s menu, add the new color and update
all menu graphs containing a vertex for $g$ and their maximum matchings.
\rapi
We select groups in decreasing order of their size and
we select colors that yield the largest gain for the group.

Observe that this method is operates similarly to the {\sl few colors} method discussed
in the previous section. The key difference is that while colors are assigned to a group's
menu, they are not rigidly assigned to the group's edges, allowing a little more
flexibility in the choice of the final edge colors.

\begin{figure}[t]
\centerline{\includegraphics[width=4in]{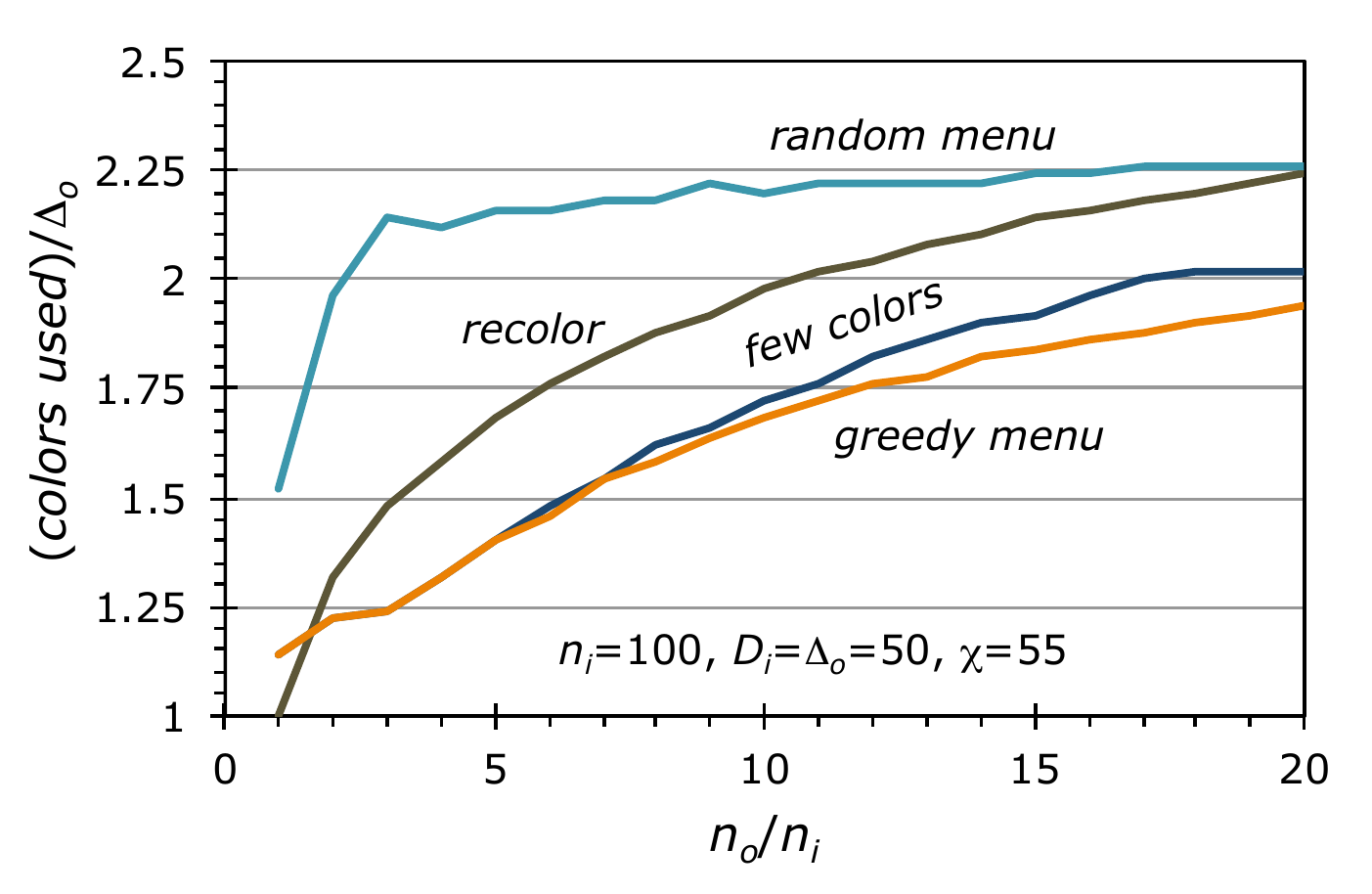}}
\caption{Effect of input/output asymmetry on the menu methods}
\label{menuAsymmetry}
\end{figure}

Figure~\ref{menuAsymmetry} shows how the two menu-based methods compare
to the {\sl recolor} and {\sl few colors} methods.
We observe that the {\sl greedy menu} method has the best performance overall,
although the {\sl few colors} method is very close for smaller asymmetries.
For random menus, the performance ratio increases very slowly
for asymmetries of two or more,. Indeed, for larger asymmetries than are shown
on the chart, it out-performs both the {\sl recolor} and {\sl few colors} methods.
Specifically, when the asymmetry is 100, the performance ratio for the
{\sl recolor} method is 2.82, for the 
{\sl few colors} method it is 2.72, for the {\sl random menu} method it is 2.44
and for the {\sl greedy menu} method it is 2.1.

\begin{figure}[t]
\centerline{\includegraphics[width=4in]{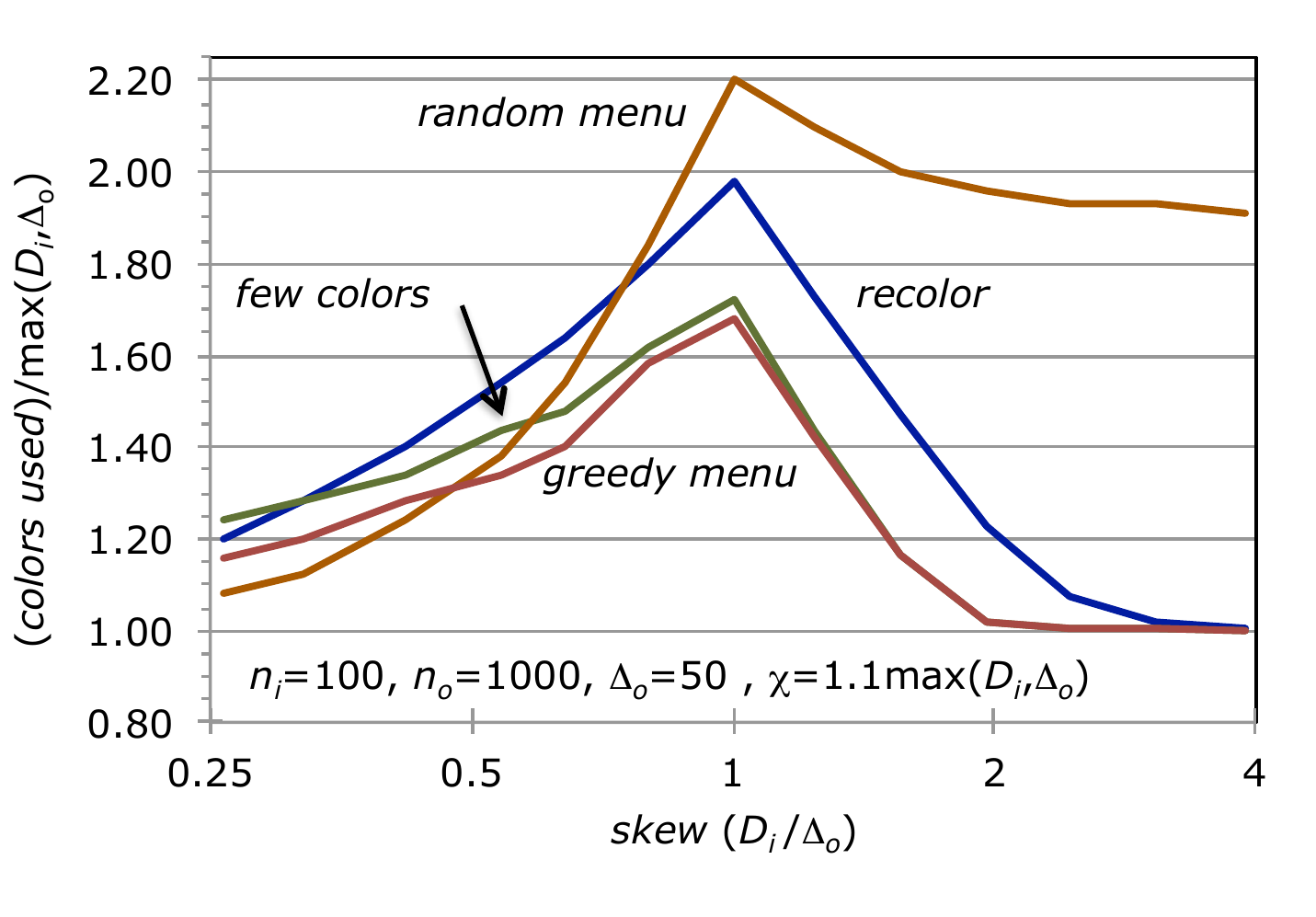}}
\caption{Effect of skew on the menu methods}
\label{menuSkew}
\end{figure}

Figure~\ref{menuSkew} shows how the performance ratio varies with skew.
For the most part, these results are consisent with what we have seen before.
However, the behavior of the {\sl random menu} method does seem a little unusual.
Its superior performance at small values of skew may reflect the fact that
the {\sl random menu} method assigns roughly the same number of colors to all
groups, providing lots of flexiblity when coloring edges, especially when $D_i$ is small.
In contrast, the {\sl greedy menu} method assigns just one or two colors to
some groups, in order to leave more available for groups that require greater
flexibility. This is advantageous when the number of colors per group is very limited,
but may not help when $D_i$ is small and there is no need to limit any group to just
one or two colors. 
For larger values of skew, the {\sl random menu} method underperforms
the others by a large margin. Apparently, with random menus, it's difficult to do
much better two colors per group, while the other methods are able to use just
one color for most groups, since they choose the colors in a more informed way.

\begin{figure}[t]
\centerline{\includegraphics[width=4in]{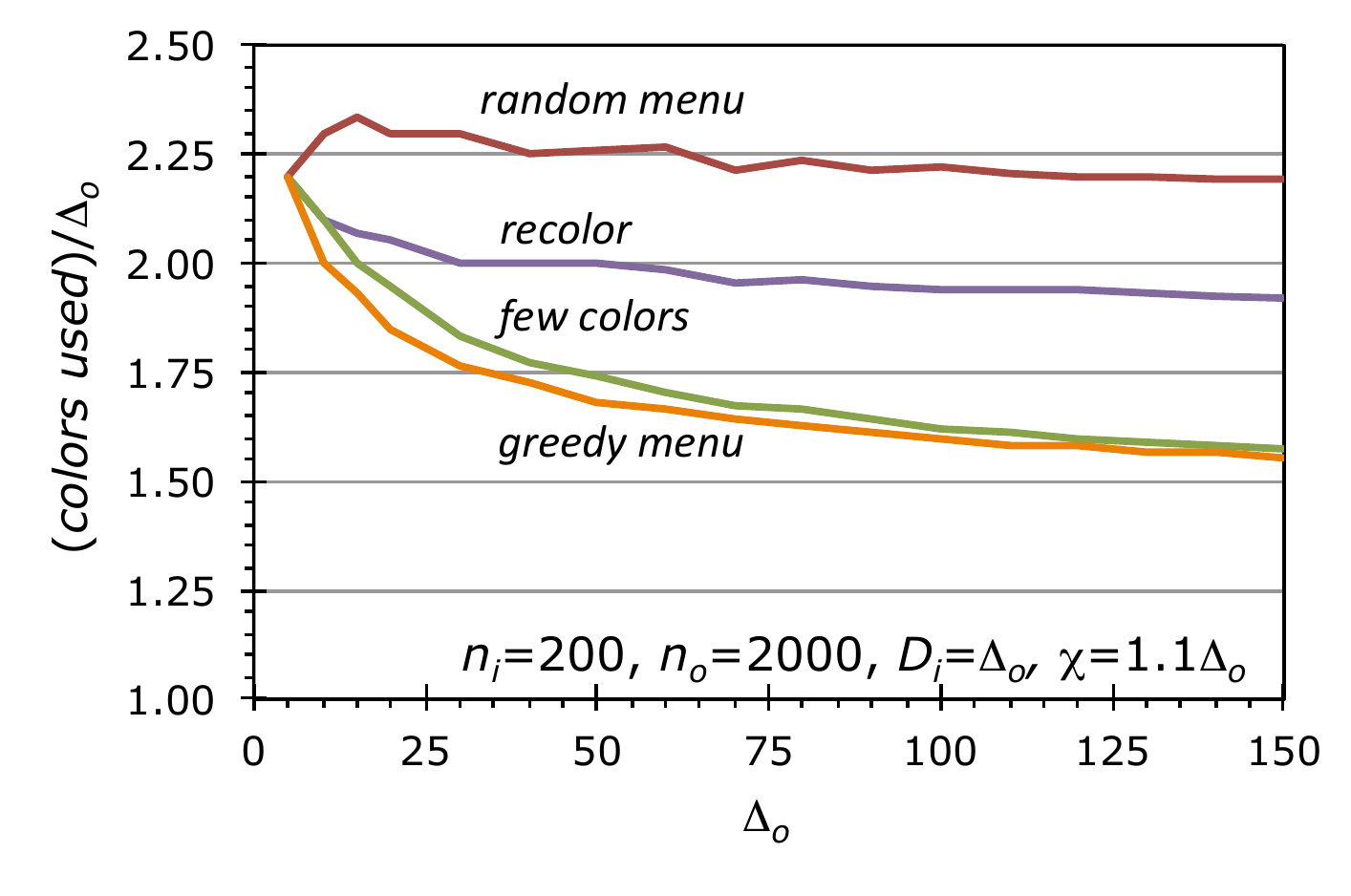}}
\caption{Effect of density on the menu methods}
\label{menuDensity}
\end{figure}

We conclude this section by examining how the performance of the {\sl few colors}
and menu methods varies with graph density
(Figure~\ref{menuDensity}).
The relative ordering of the methods remains consistent with what we have
seen before. As with the layering methods, we observe that the performance ratio
improves  as the density increases. Since dense graphs have larger values of
$\Delta_o$, the algorithms have more colors to choose from and this added
flexibility allows them to get a bit closer to the lower bound.

\section{Closing Remarks}
The edge group coloring problem is a natural generalization of the classical
edge coloring problem in graphs. While all the algorithms studied here perform well
on random graphs, only the {\sl few colors} method and {\sl random menu} method
have good analytical bounds on their performance. It seems likely that similar
bounds might be possible for the {\sl thin layers} and {\sl min color} methods,
but currently our only bound is the trivial one of $D_i\Delta_o$.
It remains open whether there exists an approximation algorithm with a
constant performance ratio.
The case of non-bipartite graphs appears to be entirely unexplored.

The menu-based methods can be applied directly to the problem of routing multicast
connections in a three stage Clos network. They could also be applied to the problem
of routing connections in an online manner. They do require the ability to rearrange
existing connections, when used in this application, but the impact of the rearrangement
on existing connections would be fairly limited.

Our results apply to the version of the multicast packet scheduling problem in which
the objective is to transfer a set of packets from inputs to outputs in a minimum
amount of time. It does not apply directly to the more practically interesting problem
of work-conserving scheduling. However, it might be possible to adapt the menu
methods to this problem. The main challenge here is that multicast packets may
involve copies going to both lightly-loaded outputs and heavily-loaded outputs.
These outputs impose conflicting requirements on the scheduler, making it difficult
to achieve strict work-conservation. However, the flexibility inherent in the menu methods
may allow for some approximate form of work-conservation.

\end{document}